\documentclass[letterpaper]{jpconf}
\usepackage{bm}
\usepackage{amssymb}
\usepackage{fix-cm}
\usepackage{graphicx}
\usepackage{hyperref}
\usepackage{wrapfig}
\usepackage[numbers,square,sort&compress]{natbib}
\newcommand{\alphas}{\ensuremath{\alpha_{\rm s}}}
\newcommand{\mup}{\ensuremath{m_{\rm u}}}
\newcommand{\mdn}{\ensuremath{m_{\rm d}}}
\newcommand{\mstr}{\ensuremath{m_{\rm s}}}
\newcommand{\mch}{\ensuremath{m_{\rm c}}}
\newcommand{\mbt}{\ensuremath{m_{\rm b}}}
\newcommand{\mtop}{\ensuremath{m_{\rm t}}}
\newcommand{\Mup}{\ensuremath{\bar{m}_{\rm u}}}
\newcommand{\Mdn}{\ensuremath{\bar{m}_{\rm d}}}
\newcommand{\Mstr}{\ensuremath{\bar{m}_{\rm s}}}
\newcommand{\Mch}{\ensuremath{\bar{m}_{\rm c}}}
\newcommand{\Mbt}{\ensuremath{\bar{m}_{\rm b}}}

\newcommand{\canada}{\includegraphics[width=9pt]{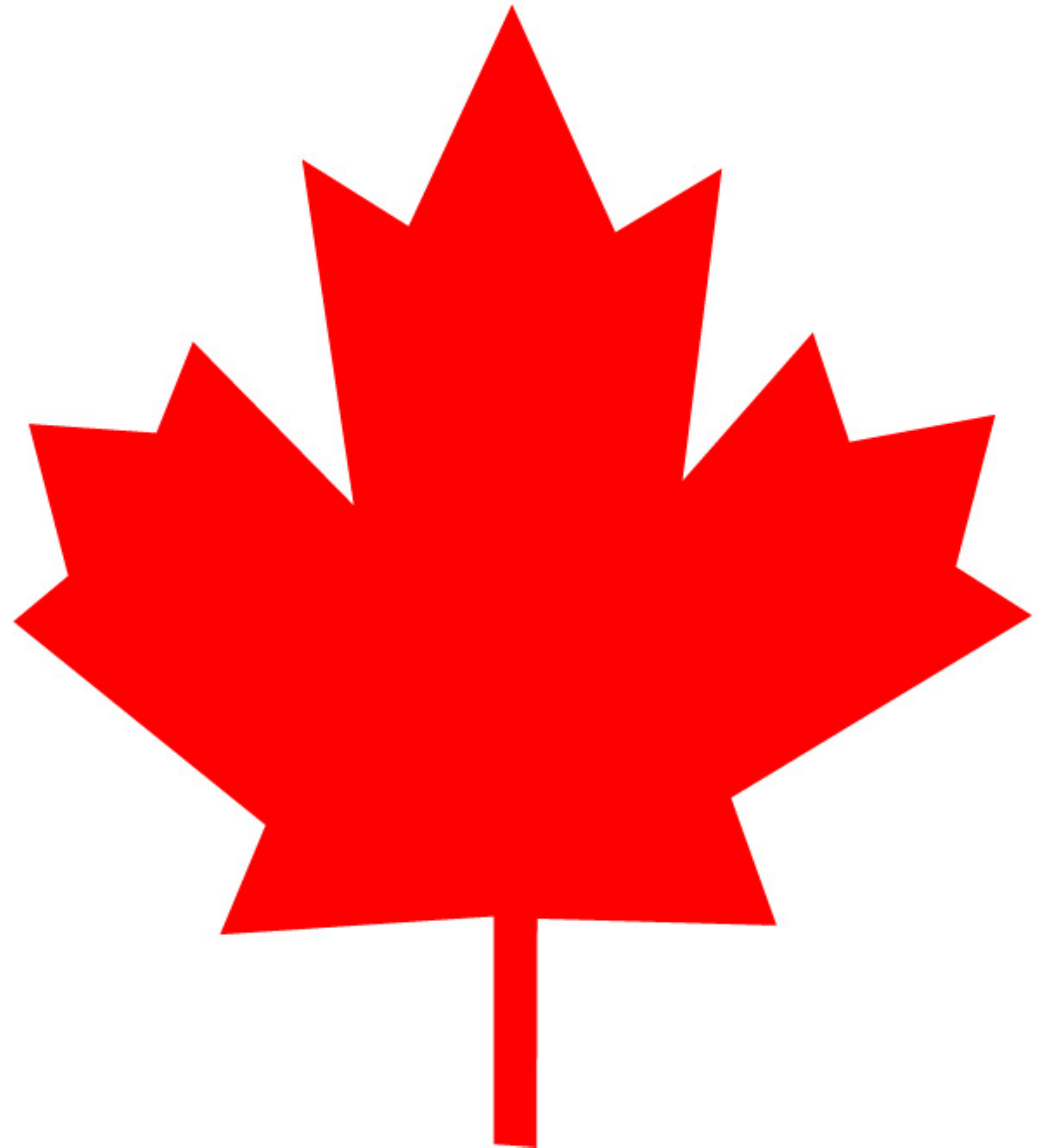}}
\begin{document}

\title{Twenty-five Years of Lattice Gauge Theory: \\
\hfill    Consequences of the QCD Lagrangian}

\author{Andreas S. Kronfeld}
\address{Theoretical Physics Department,
Fermi National Accelerator Laboratory,
Batavia, Illinois}

\begin{abstract}
When the Lake Louise Winter Institute started twenty-five years ago,
many properties of quantum chromodynamics (QCD) were \emph{believed} to 
be true, but had not been \emph{demonstrated} to be true.
This talk surveys a variety of results that have been established with 
lattice gauge theory, directly from the QCD Lagrangian, shedding light 
on the origin of (your) mass and its interplay with dynamical symmetry 
breaking, as well as some  
further intriguing features of the natural world.
\end{abstract}

\bibliographystyle{iopart-num}

\section{Solving QCD}

Quantum chromodynamics (QCD) is the modern theory of the strong nuclear 
force.
It is part of the Standard Model of elementary particles, yet also has 
profound influence on nuclear physics and on astrophysics.
It is also rich and fascinating.
In this talk, I aim to cover some results that are interesting 
in their own right, influential in a wider arena, quantitatively 
impressive, and/or qualtitatively noteworthy.

The Lagrangian of QCD has ``$1+n_f+1$'' free parameters:
\begin{equation}
    \mathcal{L}_{\mathrm{QCD}} = 
        \frac{1}{2g^2} \tr[F_{\mu\nu}F^{\mu\nu}] -
        \sum_{f=1}^{n_f} \bar{\psi}_f(/ \kern-0.65em D + m_f) \psi_f +
        \frac{i\bar{\theta}}{32\pi^2}\varepsilon^{\mu\nu\rho\sigma}
            \tr[F_{\mu\nu}F_{\rho\sigma}],
    \label{eq:lagrangian}
\end{equation}
where $F^{\mu\nu}$ is the gluon's field strength, 
$/ \kern-0.65em D=\gamma_\mu(\partial^\mu+A^\mu)$,
and $\psi_f$ denotes the quark field of flavor~$f$.
The first parameter is the gauge coupling~$g^2$, 
the next $n_f$ are the quark masses~$m_f$,
and the last, $\bar\theta$, multiplies an interaction that violates  
CP symmetry.
There are six quarks (that we know about), but at energies below the 
top, bottom, and charm thresholds, it is convenient and customary to 
absorb the short-distance effects of these quarks into a shift of $g^2$ 
and then take QCD with $n_f=5$, 4, or~3.
In addition to these shifts, the coupling $g^2$ diminishes gradually 
with increasing energy, stemming from virtual processes of gluons and 
the $n_f$ active quarks; this is called ``asymptotic freedom''~%
\cite{Politzer:1973fx,Gross:1973id}.
More generally, one could imagine a matrix in the mass term, 
$\bar{\psi}_am_{ab}\psi_b$, with eigenvalues $m_f$ in 
Eq.~(\ref{eq:lagrangian}).
In this context, the coupling multiplying 
$\varepsilon^{\mu\nu\rho\sigma}\tr[F_{\mu\nu}F_{\rho\sigma}]$ is altered:
$\bar{\theta}=\theta-\arg\det m$.
In the Standard Model, $\theta$ is considered purely chromodynamic, 
whereas $m_{ab}$ arises from Yukawa couplings between quarks and the 
weak-isodoublet Higgs boson.
Only the difference $\bar{\theta}$ is observable.

Before saying that a mathematical theory describes or explains the 
natural world, one must fix the free parameters with the corresponding 
number of measurements, in this case $1+n_f+1$.
Because the color of quarks and gluons is confined, the free parameters 
of QCD must be connected to properties of QCD's eigenstates, which are 
the bound states called hadrons.
At high energies, a sum over many hadronic states can be related to a 
sum over many quark-antiquark-gluon states (``quark-hadron duality''
\cite{Poggio:1975af}).
At lower energies, this is not possible.
In this nonperturbative regime, it is preferable to relate all the 
parameters to quantities like hadron masses, whose experimental 
interpretation is clear.
Then the obstacle is (merely) to compute the relationship between the 
QCD Lagrangian and hadronic properties.

A long-promising and now-successful approach to such computations is to 
formulate QCD as a lattice gauge theory~\cite{Wilson:1974sk}.
Then the ultraviolet cutoff (needed in any calculation) is 
built in from the outset, and the correlation functions of QCD are
mathematically well-defined.
The parameters are fixed as follows.
The electric-dipole moment of the neutron is unobservably small, 
leading to a bound $\bar{\theta}<10^{-11}$.
Such delicate cancellation of $\theta$ and $\arg\det m$ is a mystery, 
known as the strong CP problem~\cite{Kim:2008hd}, but for QCD 
calculations it simply means we can set $\bar{\theta}=0$ with no 
important consequences.
The rest are tuned to reproduce $1+n_f$ specific hadronic properties.
Even the dimensionless gauge coupling is related to a dimensionful,
measurable quantity, because the gauge coupling runs.
Thus, to lend a physical interpretation to $g^2$ one has to compute the 
energy at which the coupling reaches a fiducial value, like $g^2=1$.
But the calculations don't know \emph{a priori} about kg or~GeV/$c^2$;
instead the energy at which $g^2=1$ can be computed only relative to 
some other standard mass, such as the mass of the nucleon.%
\footnote{In principle, the nucleon mass is a good example, but it 
depends sensitively enough on quark masses that, in practice, one uses 
strange baryon masses, quarkonium mass splittings, or other quantities 
that are not as sensitive.}

It is worth solving QCD quantitatively for many practical reasons in 
particle physics, nuclear physics, and astrophysics.
In such applications, the enterprise is unabashedly pragmatic, 
with a focus on a clear understanding of uncertainties remaining 
in the calculation.
This is important but a secondary focus of this talk.

The primary focus is as follows.
The conception of QCD is rightly hailed as a triumph of reductionism, 
melding the quark model, the idea of color, and the parton model into a 
dynamical quantum field theory.
The application of QCD is, however, rich in emergent phenomena.
Symmetries emerge in idealized limits:
C, P, and T are exact when $\bar\theta=0$; chiral symmetries emerge 
when two or more quark masses vanish~\cite{Nambu:1960xd}; and 
heavy-quark symmetries are revealed as a quark mass goes to 
infinity~\cite{Shifman:1986sm,Isgur:1989vq}\canada.
More remarkable still are the dynamical phenomena that emerge, starting 
with~$\Lambda_{\rm QCD}$, the ``typical scale of QCD,'' which is not an 
input.
Much of what is ``known'' about QCD in this essentially nonperturbative 
arena has been, for a long time, based on belief:
Evidence from high-energy scattering led to the opinion that QCD
explains all of the strong interactions.
This opinion led to the belief that QCD exhibits certain properties, 
because otherwise it would not be consistent with observations.
These emergent phenomena---such as chiral symmetry breaking, the 
generation of large hadron masses despite very small quark masses, 
and the thermodynamic phase structure---are the most profound phenomena 
of gauge theories.
The aim of this talk is to survey how lattice QCD has filled many of 
these gaps, replacing belief with knowledge.

The rest of this talk is organized as follows.
Section~\ref{sec:lgt} give a short review of lattice-QCD methodology 
and jargon.
Hadron masses and their connection to chiral symmetry are discussed in 
Secs.~\ref{sec:spectrum} and~\ref{sec:XSB}.
An output of these calculations are the (in some cases remarkably 
small) quark masses and the gauge coupling; these results are discussed 
in Sec.~\ref{sec:SM}, along with a few ``tense'' results on flavor physics.
The phase structure of QCD is discussed in Sec.~\ref{sec:thermo}.
Section~\ref{sec:sum} offers some perspective.
The developments reported here took place during quarter-century 
history of the Lake Louise Winter Institute.
Many of the leading players have been Canadian, and wherever I know of
a Canadian connection I've marked the contribution with a maple 
leaf~\canada.

\section{Lattice Gauge Theory}
\label{sec:lgt}

Lattice gauge theory~\cite{Wilson:1974sk} was invented in an attempt to 
understand asymptotic freedom without introducing gauge-fixing and 
ghosts~\cite{Wilson:2004de}.
The key innovation of Ref.~\cite{Wilson:1974sk} is to formulate 
non-Abelian gauge invariance on a spacetime lattice.
Then the functional integrals defining QCD correlation functions are 
well-defined:
\begin{equation}
    \langle\bullet\rangle = \frac{1}{Z}\int \mathcal{D}A
        \mathcal{D}\psi\mathcal{D}\bar{\psi}
        \, [\bullet] \exp\left(-S\right),
    \label{eq:Z}
\end{equation}
because the measures $\mathcal{D}U$, $\mathcal{D}\psi$, 
$\mathcal{D}\bar{\psi}$ are products of a countable number of normal 
differentials. 
Here $S=\int d^4x\mathcal{L}$ is the action, $\bullet$ is just about 
anything, and $Z$ ensures $\langle1\rangle=1$.
This formulation is formally equivalent to classical statistical 
mechanics, enabling theorists to apply a larger tool-kit to 
quantum field theory.
For example, Wilson used a strong-coupling expansion to lowest order in 
$1/g^2$ to demonstrate confinement~\cite{Wilson:1974sk}.

Among the techniques of statistical mechanics, the one that has become 
an industry is to integrate expressions of the form~(\ref{eq:Z}) 
on big computers with Monte Carlo methods. 
Although lattice gauge theory defines QCD mathematically and, thus, in 
principle provides an algorithm for computing anything, as with any 
numerical analysis, compromises are necessary in practice.
To evaluate anything meaningful within a human lifetime, the integrals 
are defined at imaginary time, $t=-ix_4$, turning Feynman's phase 
factor into the damped exponential of Eq.~(\ref{eq:Z}).
A computer, obviously, has finite memory and processing power, so the 
spatial volume and time extent of the lattice are finite.

From this expression, it is straightforward to derive some simple 
results for correlation functions.
The two-point function
\begin{equation}
    \langle\pi(t)\pi^\dagger(0)\rangle = \sum_n 
        |\langle0|\hat{\pi}|\pi_n\rangle|^2\exp(-m_{\pi_n}t),
    \label{eq:2pt-m}
\end{equation}
where $\pi$ is a composite field of definite quantum numbers (e.g., of 
the pion), and the sum ranges over all radial excitations.
For time separation $t$ large enough, a fit to an exponential yields 
the lowest-lying $m_{\pi_1}$ and $|\langle0|\hat{\pi}|\pi_1\rangle|$.
This is how we compute masses.
For a transition with no hadrons in the final state, as in leptonic 
decays, simply replace $\pi(t)$ with a current~$J$:
\begin{equation}
    \langle\pi(t)\pi^\dagger(0)\rangle = \sum_n 
        \langle0|\hat{J}|\pi_n\rangle
        \langle\pi_n|\hat{\pi}^\dagger|0\rangle \exp(-m_{\pi_n}t),
    \label{eq:2pt-f}
\end{equation}
in which the only new information is $\langle0|\hat{J}|\pi_n\rangle$, 
yileding for large $t$ the matrix element of the lowest-lying state.
This is how we compute decay constants.
For a transition with one hadron in the final state, one needs a 
three-point function:
\begin{equation}
    \langle\pi(t)J(u)B^\dagger(0)\rangle = \sum_{mn} 
        \langle0|\hat{\pi}|\pi_m\rangle
        \langle\pi_n|\hat{J}|B_m\rangle
        \langle B_m|\hat{B}^\dagger|0\rangle 
        \exp[-m_{\pi_n}(t-u)-m_{B_m}u]
    \label{eq:3pt}
\end{equation}
in which the only new information is $\langle\pi_n|\hat{J}|B_m\rangle$,
yileding for large $t$ and $u$ the matrix element between the 
lowest-lying states.
This is how we compute form factors and neutral hadron oscillation 
properties.

Equations~(\ref{eq:2pt-m})--(\ref{eq:3pt}) are derived by inserting 
complete sets of eigenstates the QCD Hamiltonian.
Thus, every successful fit of these formulae for hadronic correlators 
provides \emph{a~posteriori}
incre\-mental evidence that hadrons are indeed the eigenstates of QCD.

In all cases of interest, the fermion action is of the form 
$\bar{\psi}M\psi$, where the matrix $M$ is a discretization of the 
Dirac operator (plus quark mass).
Then the fermionic integration in Eq.~(\ref{eq:Z}) can be carried out 
by hand:
\begin{equation}
    \langle\bullet\rangle = \frac{1}{Z}\int \mathcal{D}A \, [\bullet']
        \det M \exp\left(-S_{\mathrm{gauge}}\right),
    \label{eq:Zq}
\end{equation}
in which the fermionic integration replaces $\psi_i\bar{\psi}_j$ with 
$[M^{-1}]_{ij}$ to yield $\bullet'$.
Importance sampling, which is crucial, becomes possible if 
$\det M \exp\left(-S_{\mathrm{gauge}}\right)$ is positive.
In most cases, a notable exception being the case of nonzero baryon 
chemical potential, this condition holds.
\pagebreak

The determinant $\det M$ is the mathematical representation of virtual 
quark-antiquark pairs, also called sea quarks.
The matrix inverse $M^{-1}$ is the propagator of a valence quark moving 
through a stew of gluons $A$ and sea quarks $\det M$.
Several quark propagators are sewn together to form hadronic 
correlation functions, from which masses and transition matrix elements 
can be computed.
Computationally, $\det M$ is the biggest, and $M^{-1}$ the second 
biggest, challenge in lattice QCD.
The numerical algorithms become even more demanding as the quark mass 
is reduced, so in practice light quark masses are usually 2--5 times 
larger than the physical up and down quark masses.

Because $\det M$ is so CPU-intensive, for many years lattice-QCD 
calculations were carried out in the ``quenched approximation,''
in which $\det M$ is set to $1$.
Except for pilot studies of specialized new methods, the quenched 
approximation is now obsolete.
It lives on in the jargon ``unquenched lattice QCD,'' which means 
simply to do the right thing, namely compute $\det M$.
It also lives on in the jargon ``partially quenched'' QCD, which refers 
to unphysical set-ups in which $M_{\mathrm{valence}}$ and 
$M_{\mathrm{sea}}$ differ.
(Often just the value of the quark mass differs.)
This is useful because we believe we have a theory, a version of chiral 
perturbation theory, to incorporate the unphysical results into fits 
that, in the end, yield physical results~\cite{Sharpe:2000bc}.

How do the compromises of numerical lattice QCD impair the results 
discussed below? 
The imaginary time imposes no problem whatsoever for static quantities.
The finite volume introduces errors that are exponentially suppressed 
and, hence, a minor source of uncertainty.
Likewise for the finite time extent, except in thermodynamics 
(Sec.~\ref{sec:thermo}), where it becomes a tool.
Finally the unphysical light quarks are exptrapolated away with
self-consistent formulae from chiral perturbation 
theory~\cite{Bijnens:2007yd}: a physical way to think about this step 
is that we remove the cloud of unphysically massive pions and replace 
them with the real thing (to some order in chiral perturbation theory).
As in any industry, the techniques that work well for many things do 
not work for everything.

By 2003 these techniques---including a realistic formulation of 
$\det M$, had matured---making possible several 
postdictions~\cite{Davies:2003ik}\canada.
The next step was to make several \emph{pre}dictions of hadronic 
properties that were not yet, but were soon to be, measured in 
experiments.
These included form factors of $D\to Kl\nu$ and $D\to\pi l\nu$ 
decays~\cite{Aubin:2004ej}, the mass of the $B_c$ 
meson~\cite{Allison:2004be}, and the decay constants of charmed 
mesons~\cite{Aubin:2005ar}.
For a description of these developments, see 
Ref.~\cite{Kronfeld:2006sk}.

\section{Hadron Spectrum}
\label{sec:spectrum}

We compute the masses of hadrons not only ``because they are there,'' 
but also because it is interesting to see whether, and how, QCD 
generates mass. 
Particularly in extensions of the electroweak part of the Standard 
Model, there are many theoretical ideas for how mass is generated.
Among these, however, we only know for sure that Nature makes use of 
the binding energy of gauge theories.
Thus, QCD is a prototype for a mechanism that could reappear at shorter 
distance scales.

It is instructive to start by thinking semiclassically about the 
potential energy~$V(r)$ between heavy quarks at separation~$r$, or, 
equivalently, the force $F(r)=-dV/dr$.
At short distances, $V$ and~$F$ are Coul\-om\-bic, with logarithmic 
corrections from asymptotic freedom.
At large distances, the potential grows linearly, and, correspondingly, 
the force becomes a negative constant.
An accurate picture of how this force arises is as follows.
As $r$ increases, a dipole field similar to that of 
electrodynamics forms.
Gluons, being colored, attract each other, so the chromelectric field 
lines narrow, first into a sausage and eventually into a string.
The QCD flux tube is full of energy, rising linearly with~$r$.
This energy, via $m=E/c^2$, is the origin of hadron mass.

\begin{wrapfigure}{r}{3.125in}
    \hfill
    \includegraphics[width=3in]{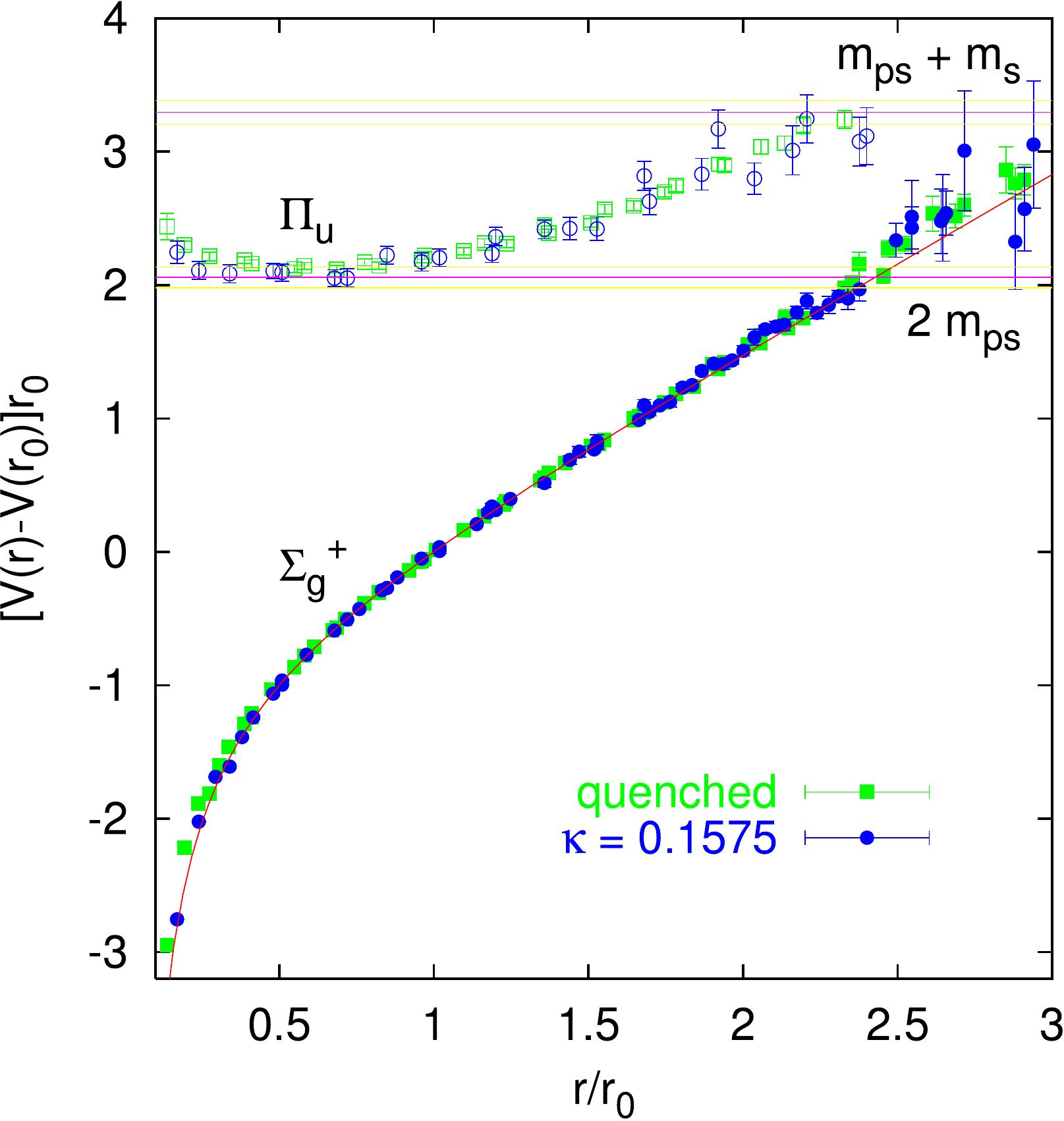}
    \caption[fig:V]{The heavy-quark potential $V(r)$ vs.~$r$, computed 
        with lattice QCD (line and points labeled $\Sigma_g^+$).
        From Ref.~\cite{Bali:2000gf}.
        (The line and points labeled $\Pi_u$ denote an excitation of 
        the interquark chromoelectric field.)
        \vspace*{12pt}}
    \label{fig:V}
\end{wrapfigure}

This picture is confirmed in detail by lattice-QCD calculations of the 
potential, as shown in Fig.~\ref{fig:V}.
In addition to establishing the Coulomb+linear behavior described above 
(the points labeled $\Sigma_g^+$), Fig.~\ref{fig:V} shows an excitation 
of the chromoelectric flux between heavy quarks 
(the points labeled $\Pi_u$).
This and further excitations elucidate how the QCD flux 
tube generates mass~\cite{Juge:2002br}\canada.

\begin{figure}[b]
    (a) \\[-2.0em]
    \begin{minipage}[b]{0.52\textwidth}
        \includegraphics[width=\textwidth]{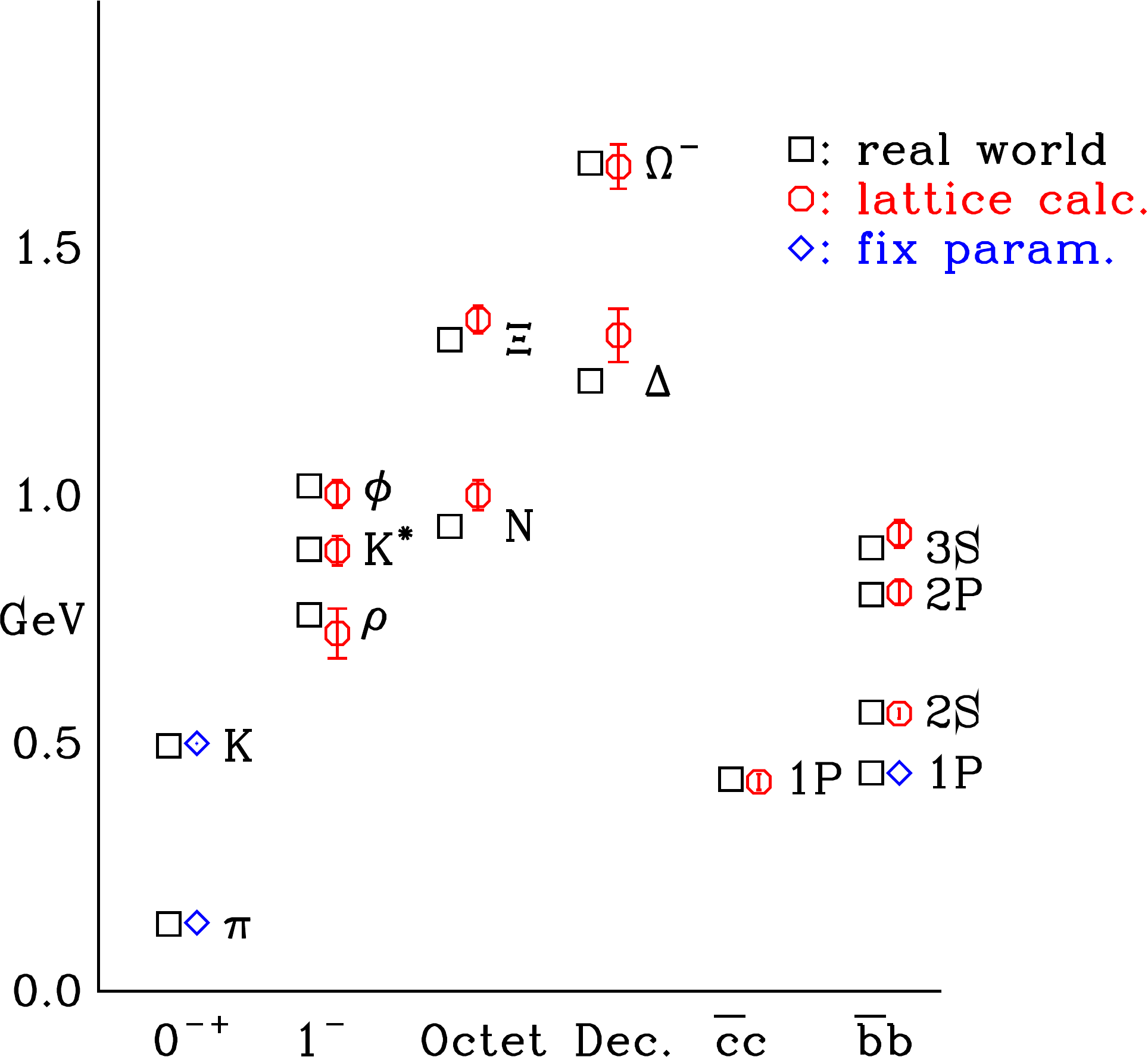}        
    \end{minipage}
    \hfill
    \begin{minipage}[b]{0.38\textwidth}
        \vspace*{-12pt}
        \hspace*{-44pt}(b)\hspace{22pt}
        \includegraphics[width=1.01\textwidth]{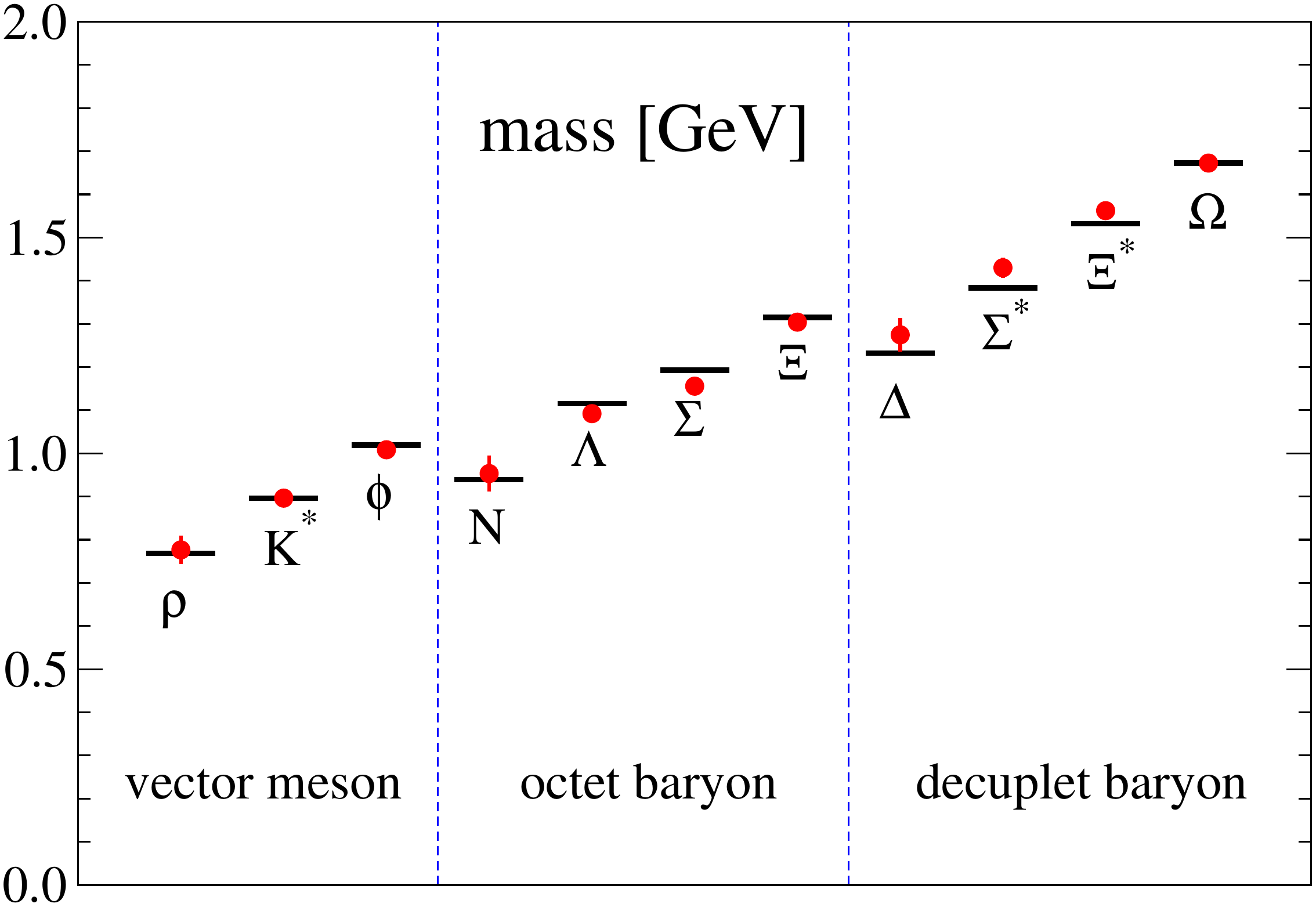}\\[12pt]
        \hspace*{-48pt}
        (c)~\includegraphics[width=1.15\textwidth]{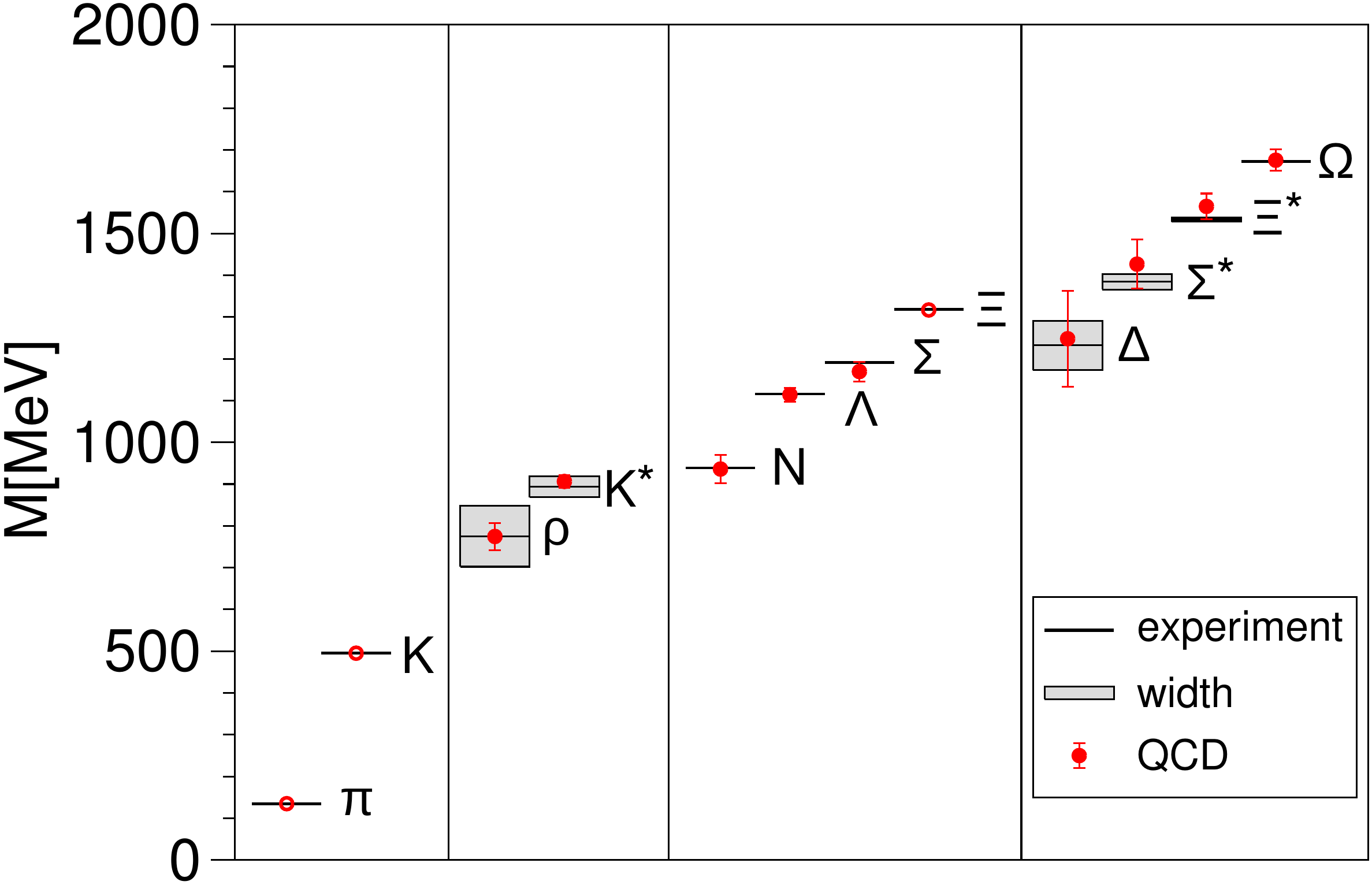}
    \end{minipage}
    \caption[fig:spectrum]{Hadron spectrum computed with lattice QCD by
        (a)~the MILC Collaboration~\cite{Aubin:2004wf,Bazavov:2009bb},
        (b)~the PACS-CS Collaboration~\cite{Aoki:2008sm}, and
        (c)~the BMW Collaboration~\cite{Durr:2008zz}.}
    \label{fig:spectrum}
\end{figure}

\vspace*{-12pt}
In addition to this appealing picture, lattice QCD has been used to
verify the mass spectrum of hadrons quantitatively, 
within a few percent. 
Figure~\ref{fig:spectrum} shows three sets of results, 
each with specific compelling features. 
Unless noted below, the error bars on the calculations encompass
all systematic uncertainties.
Figure~\ref{fig:spectrum}a shows the broadest attack on the 
spectrum \cite{Aubin:2004wf,Bazavov:2009bb}, including 
$b\bar{b}$\canada\ and $c\bar{c}$
states taken from Refs.~\cite{Gray:2005ur,Burch:2009az}.
The baryon masses are less well determined than meson masses, partly 
because of larger statistical errors, and partly because the lattice 
formulation of light quarks in Refs.~\cite{Aubin:2004wf,Bazavov:2009bb} 
is sub-optimal for baryons.
Figure~\ref{fig:spectrum}b shows the spectrum with light 
in-simulation quark masses nearly as small as $\frac{1}{2}(\mup+\mdn)$
\cite{Aoki:2008sm}.
Such small quark masses are at the frontier, and the plot omits the 
error bar associated with the continuum limit.
Finally, Fig.~\ref{fig:spectrum}c shows a complete calculation with 
good control of the baryons~\cite{Durr:2008zz}.
In particular, the nucleon mass, which provides almost all the
mass in everyday objects~\cite{Kronfeld:2008zz}, has now been 
verified within 3.5\% to arise principally from chromodynamics: 
$m=E/c^2$.

\section{Chiral Symmetry Breaking}
\label{sec:XSB}

A striking feature of the hadron spectrum is that the pion has a small 
mass, around 135--140~MeV, when most other hadrons have masses five or 
more times larger.
For example, $m_\rho=770$~MeV, $m_p=938$~MeV.
To understand the origin of the difference, Nambu~\cite{Nambu:1960xd}
applied lessons from superconductivity, noting (four years before quarks) 
that the pion's mass could be constrained to vanish by a spontaneously 
broken axial symmetry, with a small amount of explicit symmetry 
breaking allowing it to be nonzero. 

QCD explains the origin of this symmetry.
If the up and down quarks can be neglected, the Lagrangian acquires an 
$\mathrm{SU}_{\mathrm{L}}(2)\times\mathrm{SU}_{\mathrm{R}}(2)$ symmetry,
which provides a candidate axial symmetry.
The consequences of spontaneous symmetry breaking were studied further 
by Goldstone~\cite{Goldstone:1961eq}, leading to a 
formula~\cite{Goldstone:1962es},
\begin{equation}
    m_\pi^2 \langle\bar{\psi}\psi\rangle = 0,
    \label{eq:goldstone}
\end{equation}
when applied to QCD with massless up and down quarks.
The flavor-singlet expectation value $\langle\bar{\psi}\psi\rangle$ is 
called the chiral condensate.
If either factor on the left-hand side of Eq.~(\ref{eq:goldstone}) is 
nonzero, the other must vanish.

Twenty-five years ago, when the Lake Louise Winter Institute began,
most physicists were confident that QCD was a good theory of the strong 
interactions, based on, for example, its explanation of Bjorken scaling 
in deep-inelastic scattering\canada.
Because QCD was considered right, and because Nambu's picture of the 
pion was considered right, it was believed that QCD \emph{must} 
generate a chiral condensate.
There was, however, no direct calculation of 
$\langle\bar{\psi}\psi\rangle$ starting from the QCD Lagrangian, 
Eq.~(\ref{eq:lagrangian}).
Now there is.
Lattice QCD shows~\cite{Fukaya:2009fh}
\begin{equation}
    \langle\bar{\psi}\psi\rangle = 
        \left[242\pm4^{+19}_{-18}~\mathrm{MeV}\right]^3 \quad
        \mbox{($\overline{\rm MS}$ scheme at 2~GeV)},
\end{equation}
where the first uncertainty is statistical and the second a 
combination of systematics.
The quark masses have been adjusted to Nambu's idealization, 
$\mup=\mdn\to0$, $\mstr$~physical.
Spontaneous symmetry breaking also appears in two-flavor 
QCD~\cite{DeGrand:2006nv}.
The chiral condensate has been established by direct computation 
to be very significantly nonzero.
QCD breaks chiral symmetries spontaneously.

\section{SM Parameters}
\label{sec:SM}

The Standard Model has 19 free parameters, or 28 if nonzero neutrino 
masses and mixings are taken into account:
\begin{itemize}
    \item Gauge couplings: $\bm{\alphas}$, $\alpha_{\rm QED}$,
        $\alpha_{\rm W}=(M_{\rm W}/v)^2/\pi$;
    \item Lepton masses and mixing: $m_{\nu_1}$, $m_{\nu_2}$, $m_{\nu_3}$,
         $m_{\rm e}$, $m_\mu$, $m_\tau$;
         $\theta_{12}$, $\theta_{23}$, $\theta_{13}$, $\delta_{\rm PMNS}$,
         $\alpha_{21}$, $\alpha_{31}$;\footnote{For an 
             explanation of neutrino mixing parameters, see 
             Ref.~\cite{Petcov:2004wz}.}
     \item Quark masses and mixing: $\bm{\mup} e^{i\bm{\bar{\theta}}}$, 
         $\bm{\mdn}$, $\bm{\mstr}$, $\bm{\mch}$, $\bm{\mbt}$, \mtop;
         $\bm{V_{\rm us}}$, $\bm{V_{\rm cb}}$, $\bm{V_{\rm ub}}$, 
         $\bm{\delta_{\rm KM}}$;
     \item Standard electroweak symmetry breaking: $v=246$~GeV, 
         $\lambda=(M_{H}/v)^2/2$.
\end{itemize}

For ten or eleven of these parameters ($\textbf{bold}$), lattice QCD 
is either essential or important for determining their values of the 
natural world.
Lattice field theory (without QCD) is also useful for shedding light on 
the Higgs self-coupling~$\lambda$ \cite{Gerhold:2010bh} and the 
top-quark Yukawa coupling $y_{\mathrm{t}}=\sqrt{2}\mtop/v$ 
\cite{Gerhold:2009ub}, for which consistency of the field theory 
precludes arbitrary values.

\subsection{QCD Parameters}

Owing to confinement, there is no way to measure quark masses 
in a way comparable to, say, the electron mass.
Instead, a \emph{Lagrangian} definition as in 
Eq.~(\ref{eq:lagrangian}), with suitable choice of renormalization 
scheme, must be determined from measurable properties of hadrons.
For the light quarks, the simplest hadronic property is simply the 
pseudoscalar mesons masses, with the physical value obtained when the 
pion and kaon masses agree with experiment.
Three sets of results are shown in Table~\ref{tab:q}.
\begin{table}
    \centering
    \caption[tab:q]{Quark masses from lattice QCD converted to the 
    $\overline{\rm MS}$ scheme and run to the scale indicated
    (first, second, third, and fifth columns).
    Also listed are charmed and bottom masses determined from 
    $e^+e^-$ production (fourth column).}
    \label{tab:q}
    \begin{tabular}{r|*{5}{r@{$\,\pm\,$}l}}
        \hline\hline
         & 
         \multicolumn{2}{c}{Ref.~\cite{Bazavov:2009bb}} &
         \multicolumn{2}{c}{Ref.~\cite{Davies:2009ih}\canada} &
         \multicolumn{2}{c}{Ref.~\cite{Blum:2010ym}} &
         \multicolumn{2}{c}{Ref.~\cite{Chetyrkin:2009fv}} &
         \multicolumn{2}{c}{Ref.~\cite{McNeile:2010ji}\canada} \\
        \hline
        $\Mup(2~\textrm{GeV})$~[MeV] & 1.9&0.2 & 2.01&0.14 & 2.37&0.26 \\
        $\Mdn(2~\textrm{GeV})$~[MeV] & 4.6&0.3 & 4.79&0.16 & 4.52&0.30 \\
        $\Mstr(2~\textrm{GeV})$~[MeV] &  88&5   & 92.4&1.5  & 97.7&6.0  \\
        $\Mch({\bf 3}~\textrm{GeV})$~[MeV] & \multicolumn{2}{c}{} & 
            \multicolumn{2}{c}{}& \multicolumn{2}{c}{} & 986&13 & 986&10  \\
        $\Mbt({\bf10}~\textrm{GeV})$~[MeV] & \multicolumn{2}{c}{} & 
            \multicolumn{2}{c}{}& \multicolumn{2}{c}{} & 3610&16 & 3617&25 \\
        \hline\hline
    \end{tabular}
\end{table}
The results in the second column~\cite{Davies:2009ih} are derived from 
mass ratios underlying those in the first column~\cite{Bazavov:2009bb}, 
as discussed below.
The results in the third column are completely independent, in 
particular employing different methods for sea quarks and different 
approaches to electromagnetic effects.

There are two noteworthy features of these results.
First, the up and down masses are very small, about 4 and 9 times the 
tiny electron mass.
Quark masses arise from interactions with the Higgs field, 
or its surrogate in other models of eletroweak symmetry.
This sector is, thus, not the origin of much mass.
Second, $\mup$, though very small, is also very significantly far 
from zero.
This is interesting, because were $\mup=0$, then the additional 
symmetry of the Lagrangian would render $\bar{\theta}$ unphysical, 
obviating the strong CP problem.

The heavy charmed, bottom, and top quark masses are large enough that 
they can be determined with perturbative QCD from features of 
high-energy scattering cross sections and energy distributions.
For example, using perturbation theory to $\mathrm{O}(\alpha_{\rm s}^3)$ 
for  the moments in $s$ of the cross section for $e^+e^-\to c\bar{c}$, 
as a function of center-of-mass energy-squared~$s$, one finds the result 
in the fourth column, fourth row of Table~\ref{tab:q}.
The $e^+e^-$ data can be replaced with moments of the charmonium 
correlation function, 
calculated with lattice QCD~\cite{Bochkarev:1995ai,Allison:2008xk}.
Applying the same perturbative analysis yields the result in the fifth 
column, with astonishingly good agreement.
The same methods can be applied to bottom quarks, also shown in 
Table~\ref{tab:q}.

Returning to the light-quark masses, the new result of 
Ref.~\cite{Davies:2009ih} is a precise value of the (scheme independent) 
ratio $\mch/\mstr=11.85\pm0.16$.
Combining this ratio with $\bar{m}_c$~\cite{Allison:2008xk}\canada and 
the ratios $2\mstr/(\mdn+\mup)=27.3\pm0.3$ and $\mup/\mdn=0.42\pm0.04$, 
both from Ref.~\cite{Bazavov:2009bb}, leads to the values in the second 
column of Table~\ref{tab:q}.

Lattice QCD also provides excellent ways to determine the gauge 
coupling~$\alpha_{\rm s}=g^2/4\pi$.
In lattice gauge theory, the bare coupling $g_0^2$ is an input.
Alas, for most lattice gauge actions, perturbation theory in $g_0^2$ 
converges poorly~\cite{Lepage:1992xa}\canada, obstructing a 
perturbative conversion to the $\overline{\rm MS}$ or other such 
schemes.
Two other strategies are adopted to circumvent this obstacle.
One is to compute a short-distance lattice quantity---a Wilson loop, a 
Creutz ratio, or the potential at separations of order~$a$---and 
reexpress perturbation theory for the Monte Carlo results in a way that 
eliminates $g_0^2$.
The other is to compute a short-distance quantity with a continuum 
limit, and then apply continuum perturbation theory.
The quarkonium correlator used for $\mch$ and $\mbt$ is an example:
it also yields $\alpha_{\rm s}(2m_Q)$.
Other examples include the Schr\"odinger functional~\cite{Luscher:1992an} 
and the Adler function~\cite{Chetyrkin:1996cf}.

Results from several complementary lattice-QCD methods
\cite{Allison:2008xk,Davies:2008sw,Aoki:2009tf,Shintani:2010ph} are 
collected in Table~\ref{tab:alpha} and compared to an average of 
determinations from high-energy scattering and 
decays~\cite{Bethke:2009jm}.
\begin{table}
    \centering
    \caption[tab:alpha]{Values of $\alpha_{\rm s}(M_Z)$ from lattice 
        QCD and an average of determinations from high-energy scattering 
        and decays.
        A~recent update to the values on the first two 
        rows~\cite{Davies:2008sw,Allison:2008xk} can be 
        found in Ref.~\cite{McNeile:2010ji}.
        The central values and error bars from 
        Refs.~\cite{Aoki:2009tf,Shintani:2010ph} have been symmetrized 
        to ease comparison.}
    \label{tab:alpha}
    \begin{tabular}{r@{$\,\pm\,$}llll}
        \hline\hline
        \multicolumn{2}{c}{$\alpha_{\rm s}(M_Z)$} & Observable & 
            Sea formulation  & Reference \\
        \hline
        0.1183&0.0008 & Wilson loops, Creutz ratios, etc.\ & 
            2+1 asqtad staggered & HPQCD~\cite{Davies:2008sw}\canada \\
        0.1174&0.0012 & charmonium correlator & 
            2+1 asqtad staggered & HPQCD \cite{Allison:2008xk}\canada \\
        0.1197&0.0013 & Schr\"odinger functional & 
            2+1 improved Wilson & PACS-CS \cite{Aoki:2009tf} \\
        0.1185&0.0009 & Adler function & 
            2+1 overlap & JLQCD \cite{Shintani:2010ph} \\
        \hline
        0.1186&0.0011 & scattering, $\tau$ decay, etc.\ & 
            2+1(+1+1) Dirac (!) &  Bethke~\cite{Bethke:2009jm} \\
        \hline\hline
    \end{tabular}
\end{table}
One sees excellent consistency among results with different 
discretizations of the determinant for sea quarks.
An important source of uncertainty is the truncation of perturbation 
theory, including strategies for matching to the $\overline{\rm MS}$ 
scheme, and running to scale $M_Z$.
In the example of the lattice-scale loops, an independent analysis of 
the data from Ref.~\cite{Davies:2008sw} has been carried out, yielding
$\alpha_{\rm s}(M_Z)=0.1192\pm 0.011$~\cite{Maltman:2008bx}\canada,
to be compared with the first line of Table~\ref{tab:alpha}.

As mentioned above, QCD is a union of the quark model of hadrons and 
the parton model of high-energy scattering.
The agreement of the lattice-QCD results for $\alphas$, as well as for 
$\mch$ and $\mbt$, shows that hadrons and partons share the same QCD 
parameters, demonstrating QCD's breadth: 
the QCD of hadrons is the QCD of partons.

\subsection{Flavor Physics}
\label{sec:ckm}

Like the quark masses, the quark mixing, or Cabibbo-Kobayashi-Maskawa 
(CKM), matrix~\cite{Cabibbo:1963yz,Kobayashi:1973fv} arises from the 
electroweak interactions.
In the Standard Model, the masses are proportional to the eigenvalues 
of the Yukawa-coupling matrices between the quarks and the Higgs 
doublet.
The CKM matrix is the observable part of the transformations from the 
fields interacting with the weak gauge bosons to their mass eigenstates.
Symmetries of the gauge interactions make many components of these 
transformations unobervable.
For three generations, three mixing angles and one CP-violating phase 
remain to account for all the flavor- and CP-violation in nature.

Lattice QCD calculations have played a key role in many aspects of 
flavor physics.
A recent, comprehensive overview can be found in 
Ref.~\cite{Laiho:2009eu}, so here we shall simply note some examples in 
which tension between the experiments and the Standard Model have 
appeared.
These are tantalizing, because Standard CP violation seems insufficient 
to explain the baryon asymmetry of the universe.
Tension appears in the global fit to the four CKM 
parameters~\cite{Lunghi:2008aa} and also in several specific 
flavor-changing processes. 

As anticipated, lattice-QCD calculations of neutral kaon 
mixing~\cite{Antonio:2007pb,Aubin:2009jh} have improved such that the 
``standard'' Standard-Model analysis had to re-incorporate certain 
effects of a few percent.
(See Refs.~\cite{Anikeev:2001rk,Buras:2008nn} for details.)
With the re-improved formula, the tension in the global fit 
strengthens~\cite{Buras:2008nn,Lunghi:2009ke}.

Recent measurements of some purely leptonic decays 
$B^+\to\tau^+\nu$, $D_s\to\mu^+\nu$, and $D_s\to\tau^+\nu$ are somewhat 
in excess of the Standard-Model prediction for the branching ratio.
A~crucial ingredient are the decay constants $f_B$ and $f_{D_s}$ from 
lattice QCD~\cite{Aubin:2005ar,Follana:2007uv,Gray:2005ad,Bernard:2009wr}.
The nearly $2\sigma$ excess of $B^+\to\tau^+\nu$ is usually interpreted 
as a possible signal of charged Higgs 
bosons~\cite{Akeroyd:2007eh,Deschamps:2009rh}, but then the 
non-Standard amplitude has to be around $-110$\% of the Standard amplitude.
The excesses of $D_s\to l^+\nu$ could be due to 
leptoquarks~\cite{Dobrescu:2008er}, with an few-percent amplitude 
constructively interfering.
Note that the tension in this mode, which was once nearly $4\sigma$, is 
now below $2\sigma$, as improved calculations and, especially, new 
measurements have come out~\cite{Kronfeld:2009cf}.

\section{Thermodynamics}
\label{sec:thermo}

\begin{figure}[b]
    (a)\hspace*{-1.7em}
    \includegraphics[width=0.55\textwidth]{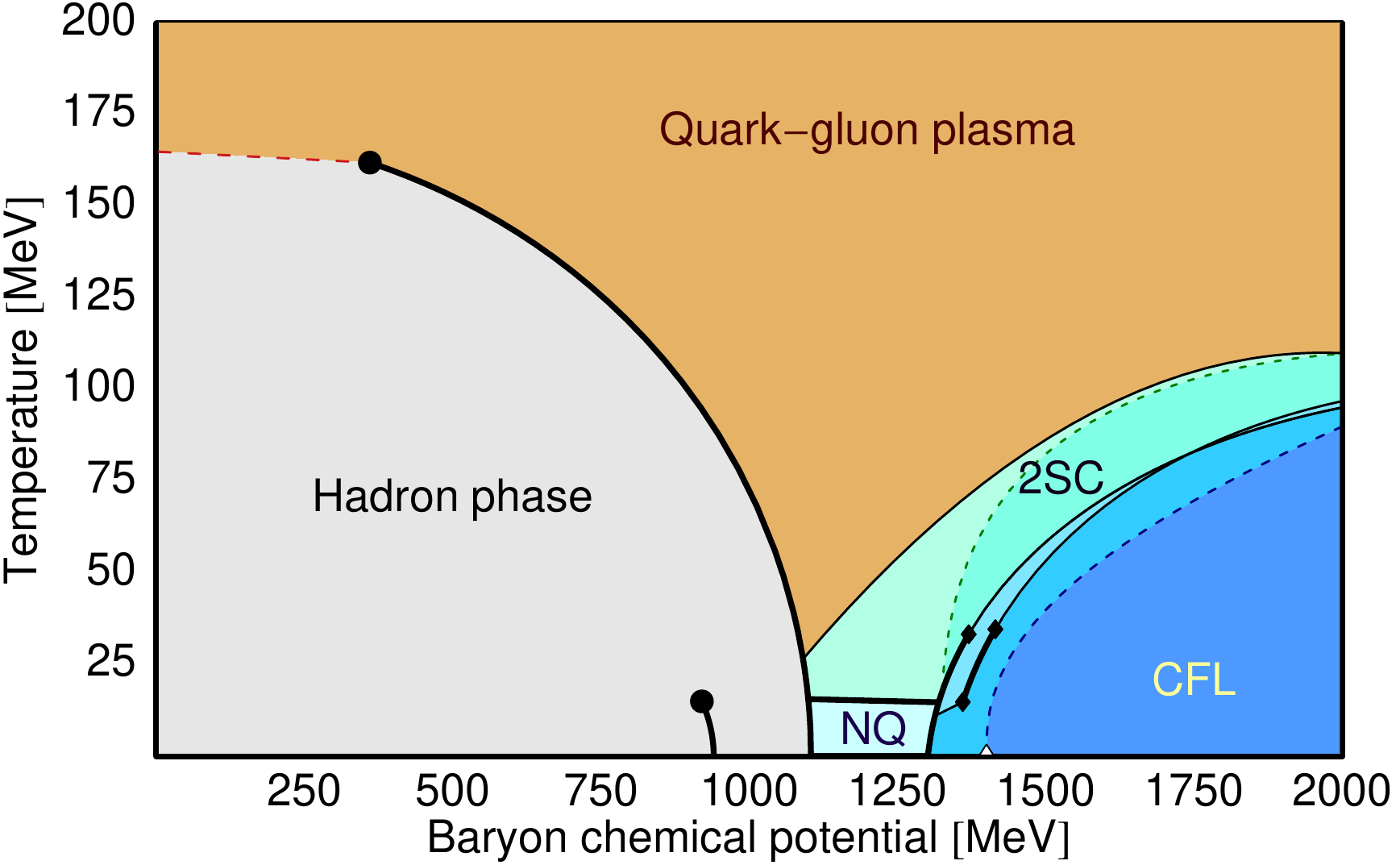}
    \hfill
    (b)\hspace*{-0.5em}\includegraphics[width=0.40\textwidth]{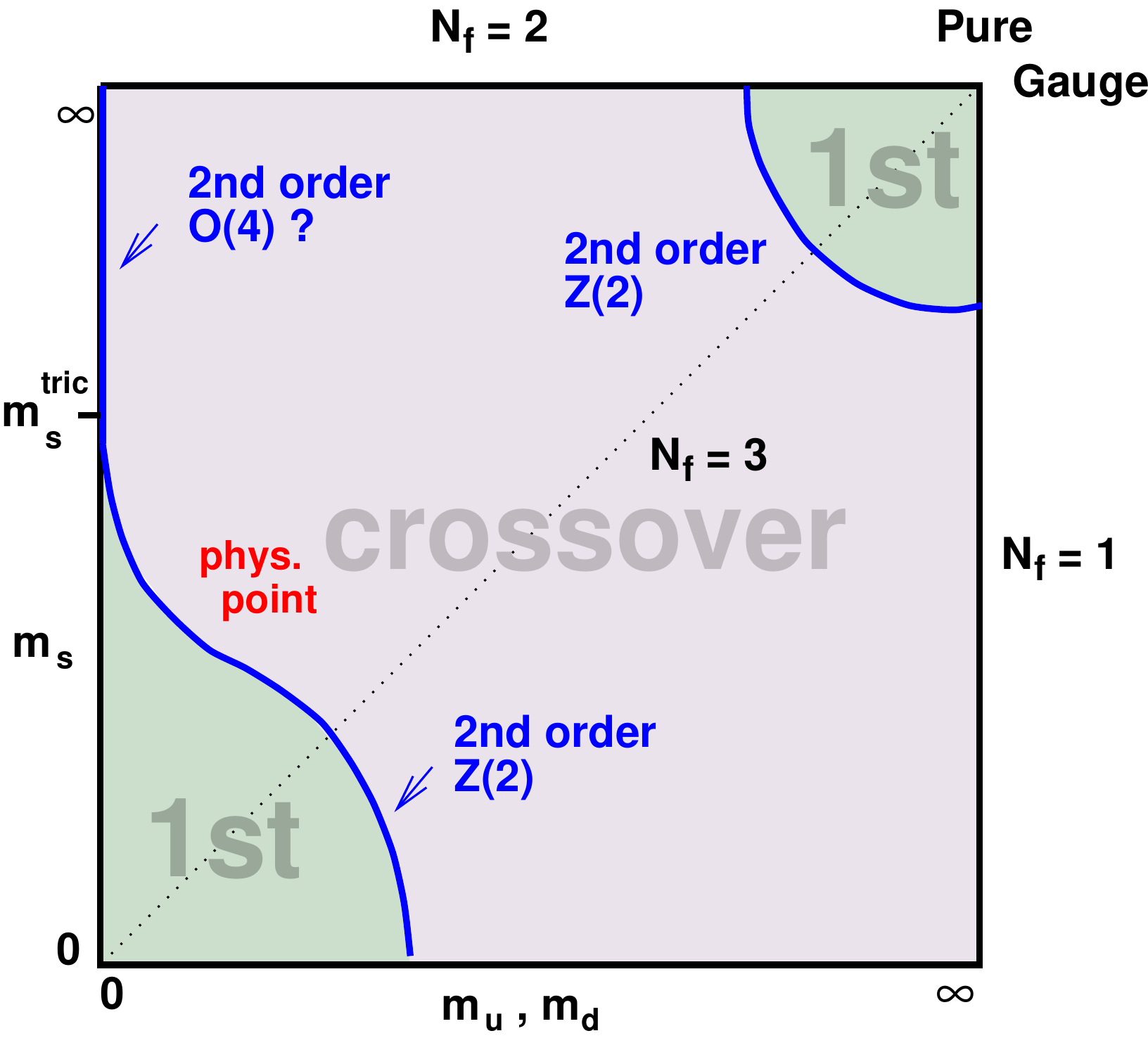}
    \caption[fig:phase]{QCD phase diagrams, (a) in the $\mu$-$T$ plane 
        (image from Ref.~\cite{Ruester:2005jc});
        (b)~at $\mu=0$, $T=T_c$, showing the order of the transition in 
        the $\mstr$-$\frac{1}{2}(\mup+\mdn)$ plane
        (image from Ref.~\cite{deForcrand:2008vr}).}
    \label{fig:phase}
\end{figure}

Like any physical system, QCD has thermodynamic properties.
During the early universe, the temperature was much hotter than it is 
now.
In neutron stars, the baryon density is much higher than in normal 
nuclear matter.
A sketch of the phase diagram is shown in Fig.~\ref{fig:phase}a, based 
on lattice-QCD studies and models~\cite{Ruester:2005jc}.
A~surprising result is that the transition between the hadronic phase 
and the ``quark-gluon plasma'' is a smooth crossover, rather than a 
first- or second-order phase transition~\cite{Aoki:2006we,Bazavov:2009zn}.
This means that as the early universe cools, the hot matter 
becomes more and more like a gas of distinct hadrons.
With a genuine phase transition, bubbles of the hadronic phase would 
form.
At nonzero baryon density (chemical potential~$\mu$), it is thought 
that the transition becomes first order, but the matter is not yet 
settled~\cite{deForcrand:2008vr}.
For further discussion, see Refs~\cite{DeTar:2009ef,Fodor:2009ax}.

It may be worth clarifying what the quark-gluon plasma is.
QCD thermodynamics is, as one would expect, based on the canonical 
ensemble, with thermal averages
\begin{equation}
    \langle\bullet\rangle = 
        \frac{\Tr\left[\bullet\,e^{-\hat{H}/T}\right]}{\Tr e^{-\hat{H}/T}},
\end{equation}
where $T$ is the temperature.
In quantum field theory formulated as in Eq.~(\ref{eq:Z}), the time 
extent $N_4$ specifies a temperature $T=(N_4a)^{-1}$.
The trace $\Tr$ is over the Hilbert space of the QCD Hamiltonian 
$\hat{H}$.
The eigenstates---aka hadrons---do not change with $T$, but as $T$ 
increases the propagation of a single source of color can change.
First, thermal fluctuations encompass states with many overlapping 
hadrons, so color can propagate from one hadron to the next, 
as if deconfined.
Second, the thermal average applies nearly equal weights to states of 
both parities, so chiral symmetry is restored---the thermal average of 
$\bar{\psi}\psi$ vanishes even if the \emph{vacuum} expectation value 
does not.
With a smooth crossover, these changes need not emerge at the 
same $T$, but, in practice, 
it seems they do~\cite{Aoki:2006we,Bazavov:2009zn}.
The picture of thermal averages over hadronic eigenstates and the 
crossover nature of the transition may help us understand why hadron 
gas models of the transition are so successful.

The nature of the QCD phase transition is influenced by the physical 
values of the light (up, down, and strange) quark masses, as sketched in
Fig.~\ref{fig:phase}b.
For vanishing quark masses, the transition would be first order.
The ratio $2\mstr/(\mup+\mdn)$ is well constrained by chiral symmetry 
(and substantiated by explicit calculation, as in Table~\ref{tab:q}).
But the masses are just large enough to push the QCD system into the 
region of crossover. 
If the light quark masses---crucially $\mstr$---were around half their 
physical size, the universe would cool through a first-order transition.
What kind of fluke is this?

\section{Summary and Challenges}
\label{sec:sum}

During the twenty-five years of the Lake Louise Winter Institute, 
lattice gauge theory has developed several ideas about QCD, nurturing 
them from ``QCD \emph{should} work this way'' to
``QCD  \emph{does}  work this way.''
Quantitative precision on $\alphas$ and heavy-quark masses reassures us 
that the QCD of hadrons is the QCD of partons.
Accurate calculations of the hadron masses and the chiral condensate 
show how QCD generates mass and that QCD breaks chiral symmetry.
The dependence of hadron masses on quark masses leads to the conclusion 
that the masses generated via interaction with the Higgs field is very 
small.
Your mass is $E/c^2$ from QCD.
Furthermore, the up quark's mass, though small, clearly does not vanish.
The nonzero up, down, and strange masses make a qualitative 
difference to the early universe, because at physical quark mass (and 
low density) the QCD phase transition is actually a smooth crossover.

On the quantitative front, QCD faces many challenges.
Hints of non-Standard processes in flavor physics require ever more 
precise calculations.
It is fairly certain that the Standard amount of CP violation is 
insufficient to explain the baryon asymmetry, so it is plausible that 
one of these hints will settle into real evidence.
The advent of the LHC calls for other calculations that still lie 
beyond today's precision frontier of lattice QCD.
For example, reliable moments of the gluon density inside the proton 
could help reduce uncertainties in LHC cross sections.
If the LHC uncovers evidence for a dynamical mechanism breaking 
electroweak symmetry (similar in some, but not all, ways to QCD), 
lattice gauge theory will be necessary for non-QCD models (and, 
eventually, a new theory) \cite{Fleming:2008gy,Pallante:2009hu}.

Nuclear physics is an arena where the need for computational lattice 
gauge theory is exploding \cite{Beane:2008dv}.
In many cases, the same basic methods 
[Eqs.~(\ref{eq:2pt-m})--(\ref{eq:3pt})] apply, 
but in others the technology has to be extended or 
invented~\cite{Bulava:2009jb}.
Nuclear lattice QCD overlaps with astrophysics:
better methods for nonzero baryon chemical potential would permit 
studies of the phases inside neutron stars, not to mention even denser 
phases shown in Fig.~\ref{fig:phase} \cite{Alford:2007xm};
meanwhile, calculations of hyperon-nucleon interactions shed light on 
strangeness in neutron stars~\cite{Beane:2006gf}.

\bibliography{lqcd}

\providecommand{\newblock}{}
\begin{thebibliography}{10}
\expandafter\ifx\csname url\endcsname\relax
  \def\url#1{{\tt #1}}\fi
\expandafter\ifx\csname urlprefix\endcsname\relax\def\urlprefix{URL }\fi
\providecommand{\eprint}[1]{\href{http://arXiv.org/abs/#1/}{#1}}

\bibitem{Politzer:1973fx}
Politzer H~D 1973 {\em Phys. Rev. Lett.\/} {\bf 30} 1346--1349

\bibitem{Gross:1973id}
Gross D~J and Wilczek F 1973 {\em Phys. Rev. Lett.\/} {\bf 30} 1343--1346

\bibitem{Poggio:1975af}
Poggio E~C, Quinn H~R and Weinberg S 1976 {\em Phys. Rev.\/} {\bf D13}
  1958--1968

\bibitem{Wilson:1974sk}
Wilson K~G 1974 {\em Phys. Rev.\/} {\bf D10} 2445--2459

\bibitem{Kim:2008hd}
Kim J~E and Carosi G 2010 {\em Rev. Mod. Phys.\/} {\bf 82} 557--601
  (\textit{Preprint} \eprint{0807.3125})

\bibitem{Nambu:1960xd}
Nambu Y 1960 {\em Phys. Rev. Lett.\/} {\bf 4} 380--382

\bibitem{Shifman:1986sm}
Shifman M~A and Voloshin M~B 1987 {\em Sov. J. Nucl. Phys.\/} {\bf 45} 292

\bibitem{Isgur:1989vq}
Isgur N and Wise M~B 1989 {\em Phys. Lett.\/} {\bf B232} 113--117

\bibitem{Wilson:2004de}
Wilson K~G 2005 {\em Nucl. Phys. Proc. Suppl.\/} {\bf 140} 3--19
  (\textit{Preprint} \eprint{hep-lat/0412043})

\bibitem{Sharpe:2000bc}
Sharpe S~R and Shoresh N 2000 {\em Phys. Rev.\/} {\bf D62} 094503
  (\textit{Preprint} \eprint{hep-lat/0006017})

\bibitem{Bijnens:2007yd}
Bijnens J 2007 {\em PoS\/} {\bf LAT2007} 004 (\textit{Preprint}
  \eprint{0708.1377})

\bibitem{Davies:2003ik}
Davies C~T~H {\em et~al.\/} (HPQCD, MILC, and Fermilab Lattice) 2004 {\em Phys.
  Rev. Lett.\/} {\bf 92} 022001 (\textit{Preprint} \eprint{hep-lat/0304004})

\bibitem{Aubin:2004ej}
Aubin C {\em et~al.\/} (Fermilab Lattice and MILC) 2005 {\em Phys. Rev.
  Lett.\/} {\bf 94} 011601 (\textit{Preprint} \eprint{hep-ph/0408306})

\bibitem{Allison:2004be}
Allison I~F {\em et~al.\/} (HPQCD and Fermilab Lattice) 2005 {\em Phys. Rev.
  Lett.\/} {\bf 94} 172001 (\textit{Preprint} \eprint{hep-lat/0411027})

\bibitem{Aubin:2005ar}
Aubin C {\em et~al.\/} (Fermilab Lattice and MILC) 2005 {\em Phys. Rev.
  Lett.\/} {\bf 95} 122002 (\textit{Preprint} \eprint{hep-lat/0506030})

\bibitem{Kronfeld:2006sk}
Kronfeld A~S (Fermilab Lattice) 2006 {\em J. Phys. Conf. Ser.\/} {\bf 46}
  147--151 (\textit{Preprint} \eprint{hep-lat/0607011})

\bibitem{Juge:2002br}
Juge K~J, Kuti J and Morningstar C 2003 {\em Phys. Rev. Lett.\/} {\bf 90}
  161601 (\textit{Preprint} \eprint{hep-lat/0207004})

\bibitem{Bali:2000gf}
Bali G~S 2001 {\em Phys. Rept.\/} {\bf 343} 1--136 (\textit{Preprint}
  \eprint{hep-ph/0001312})

\bibitem{Aubin:2004wf}
Aubin C {\em et~al.\/} (MILC) 2004 {\em Phys. Rev.\/} {\bf D70} 094505
  (\textit{Preprint} \eprint{hep-lat/0402030})

\bibitem{Bazavov:2009bb}
Bazavov A {\em et~al.\/} 2010 {\em Rev. Mod. Phys.\/} {\bf 82} 1349--1417
  (\textit{Preprint} \eprint{0903.3598})

\bibitem{Aoki:2008sm}
Aoki S {\em et~al.\/} (PACS-CS) 2009 {\em Phys. Rev.\/} {\bf D79} 034503
  (\textit{Preprint} \eprint{0807.1661})

\bibitem{Durr:2008zz}
D\"urr S {\em et~al.\/} (BMW) 2008 {\em Science\/} {\bf 322} 1224--1227
  (\textit{Preprint} \eprint{0906.3599})

\bibitem{Gray:2005ur}
Gray A {\em et~al.\/} (HPQCD) 2005 {\em Phys. Rev.\/} {\bf D72} 094507
  (\textit{Preprint} \eprint{hep-lat/0507013})

\bibitem{Burch:2009az}
Burch T {\em et~al.\/} (Fermilab Lattice and MILC) 2010 {\em Phys. Rev.\/} {\bf
  D81} 034508 (\textit{Preprint} \eprint{0912.2701})

\bibitem{Kronfeld:2008zz}
Kronfeld A~S 2008 {\em Science\/} {\bf 322} 1198--1199 (\emph{Preprint}
  \href{http://lss.fnal.gov/archive/test-fn/0000/fermilab-fn-0828-t.shtml}{FER%
MILAB-FN-0828-T})

\bibitem{Goldstone:1961eq}
Goldstone J 1961 {\em Nuovo Cim.\/} {\bf 19} 154--164

\bibitem{Goldstone:1962es}
Goldstone J, Salam A and Weinberg S 1962 {\em Phys. Rev.\/} {\bf 127} 965--970

\bibitem{Fukaya:2009fh}
Fukaya H {\em et~al.\/} (JLQCD) 2010 {\em Phys. Rev. Lett.\/} {\bf 104} 122002
  (\textit{Preprint} \eprint{0911.5555})

\bibitem{DeGrand:2006nv}
DeGrand T, Liu Z and Schaefer S 2006 {\em Phys. Rev.\/} {\bf D74} 094504
  (\textit{Preprint} \eprint{hep-lat/0608019})

\bibitem{Petcov:2004wz}
Petcov S~T 2004 {\em New J. Phys.\/} {\bf 6} 109

\bibitem{Gerhold:2010bh}
Gerhold P and Jansen K 2010 {\em JHEP\/} {\bf 1004} 094 (\textit{Preprint}
  \eprint{1002.4336})

\bibitem{Gerhold:2009ub}
Gerhold P and Jansen K 2009 {\em JHEP\/} {\bf 0907} 025 (\textit{Preprint}
  \eprint{0902.4135})

\bibitem{Davies:2009ih}
Davies C~T~H {\em et~al.\/} (HPQCD) 2010 {\em Phys. Rev. Lett.\/} {\bf 104}
  132003 (\textit{Preprint} \eprint{0910.3102})

\bibitem{Blum:2010ym}
Blum T {\em et~al.\/} 2010  (\textit{Preprint} \eprint{1006.1311})

\bibitem{Chetyrkin:2009fv}
Chetyrkin K~G {\em et~al.\/} 2009 {\em Phys. Rev.\/} {\bf D80} 074010
  (\textit{Preprint} \eprint{0907.2110})

\bibitem{McNeile:2010ji}
McNeile C, Davies C~T~H, Follana E, Hornbostel K and Lepage G~P (HPQCD) 2010
  (\textit{Preprint} \eprint{1004.4285})

\bibitem{Bochkarev:1995ai}
Bochkarev A and De~Forcrand P 1996 {\em Nucl. Phys.\/} {\bf B477} 489--520
  (\textit{Preprint} \eprint{hep-lat/9505025})

\bibitem{Allison:2008xk}
Allison I {\em et~al.\/} (HPQCD) 2008 {\em Phys. Rev.\/} {\bf D78} 054513
  (\textit{Preprint} \eprint{0805.2999})

\bibitem{Lepage:1992xa}
Lepage G~P and Mackenzie P~B 1993 {\em Phys. Rev.\/} {\bf D48} 2250--2264
  (\textit{Preprint} \eprint{hep-lat/9209022})

\bibitem{Luscher:1992an}
L\"uscher M, Narayanan R, Weisz P and Wolff U 1992 {\em Nucl. Phys.\/} {\bf
  B384} 168--228 (\textit{Preprint} \eprint{hep-lat/9207009})

\bibitem{Chetyrkin:1996cf}
Chetyrkin K~G, K\"uhn J~H and Steinhauser M 1996 {\em Nucl. Phys.\/} {\bf B482}
  213--240 (\textit{Preprint} \eprint{hep-ph/9606230})

\bibitem{Davies:2008sw}
Davies C~T~H {\em et~al.\/} (HPQCD) 2008 {\em Phys. Rev.\/} {\bf D78} 114507
  (\textit{Preprint} \eprint{0807.1687})

\bibitem{Aoki:2009tf}
Aoki S {\em et~al.\/} (PACS-CS) 2009 {\em JHEP\/} {\bf 10} 053
  (\textit{Preprint} \eprint{0906.3906})

\bibitem{Shintani:2010ph}
Shintani E {\em et~al.\/} (JLQCD) 2010  (\textit{Preprint} \eprint{1002.0371})

\bibitem{Bethke:2009jm}
Bethke S 2009 {\em Eur. Phys. J.\/} {\bf C64} 689--703 (\textit{Preprint}
  \eprint{0908.1135})

\bibitem{Maltman:2008bx}
Maltman K, Leinweber D, Moran P and Sternbeck A 2008 {\em Phys. Rev.\/} {\bf
  D78} 114504 (\textit{Preprint} \eprint{0807.2020})

\bibitem{Cabibbo:1963yz}
Cabibbo N 1963 {\em Phys. Rev. Lett.\/} {\bf 10} 531--533

\bibitem{Kobayashi:1973fv}
Kobayashi M and Maskawa T 1973 {\em Prog. Theor. Phys.\/} {\bf 49} 652--657

\bibitem{Laiho:2009eu}
Laiho J, Lunghi E and Van~de Water R~S 2010 {\em Phys. Rev.\/} {\bf D81} 034503
  (\textit{Preprint} \eprint{0910.2928})

\bibitem{Lunghi:2008aa}
Lunghi E and Soni A 2008 {\em Phys. Lett.\/} {\bf B666} 162--165
  (\textit{Preprint} \eprint{0803.4340})

\bibitem{Antonio:2007pb}
Antonio D~J {\em et~al.\/} (RBC and UKQCD) 2008 {\em Phys. Rev. Lett.\/} {\bf
  100} 032001 (\textit{Preprint} \eprint{hep-ph/0702042})

\bibitem{Aubin:2009jh}
Aubin C, Laiho J and Van~de Water R~S 2010 {\em Phys. Rev.\/} {\bf D81} 014507
  (\textit{Preprint} \eprint{0905.3947})

\bibitem{Anikeev:2001rk}
Anikeev K {\em et~al.\/} 2001 {\em $B$ Physics at the Tevatron: Run II and
  Beyond\/} (Batavia: Fermilab) (\textit{Preprint} \eprint{hep-ph/0201071})

\bibitem{Buras:2008nn}
Buras A~J and Guadagnoli D 2008 {\em Phys. Rev.\/} {\bf D78} 033005
  (\textit{Preprint} \eprint{0805.3887})

\bibitem{Lunghi:2009ke}
Lunghi E and Soni A 2010 {\em Phys. Rev. Lett.\/} {\bf 104} 251802
  (\textit{Preprint} \eprint{0912.0002})

\bibitem{Follana:2007uv}
Follana E, Davies C~T~H, Lepage G~P and Shigemitsu J (HPQCD) 2008 {\em Phys.
  Rev. Lett.\/} {\bf 100} 062002 (\textit{Preprint} \eprint{0706.1726})

\bibitem{Gray:2005ad}
Gray A {\em et~al.\/} (HPQCD) 2005 {\em Phys. Rev. Lett.\/} {\bf 95} 212001
  (\textit{Preprint} \eprint{hep-lat/0507015})

\bibitem{Bernard:2009wr}
Bernard C {\em et~al.\/} 2008 {\em PoS\/} {\bf LATTICE2008} 278
  (\textit{Preprint} \eprint{0904.1895})

\bibitem{Akeroyd:2007eh}
Akeroyd A~G and Chen C~H 2007 {\em Phys. Rev.\/} {\bf D75} 075004
  (\textit{Preprint} \eprint{hep-ph/0701078})

\bibitem{Deschamps:2009rh}
Deschamps O {\em et~al.\/} 2009  (\textit{Preprint} \eprint{0907.5135})

\bibitem{Dobrescu:2008er}
Dobrescu B~A and Kronfeld A~S 2008 {\em Phys. Rev. Lett.\/} {\bf 100} 241802
  (\textit{Preprint} \eprint{0803.0512})

\bibitem{Kronfeld:2009cf}
Kronfeld A~S 2009 {\em XXIX Physics in Collision\/} ed Kawagoe K {\em et~al.\/}
  (Tokyo: Universal Academy Press) (\textit{Preprint} \eprint{0912.0543})

\bibitem{Ruester:2005jc}
R\"uster S~B, Werth V, Buballa M, Shovkovy I~A and Rischke D~H 2005 {\em Phys.
  Rev.\/} {\bf D72} 034004 (\textit{Preprint} \eprint{hep-ph/0503184})

\bibitem{deForcrand:2008vr}
de~Forcrand P and Philipsen O 2008 {\em JHEP\/} {\bf 11} 012 (\textit{Preprint}
  \eprint{0808.1096})

\bibitem{Aoki:2006we}
Aoki Y, Endrodi G, Fodor Z, Katz S~D and Szabo K~K 2006 {\em Nature\/} {\bf
  443} 675--678 (\textit{Preprint} \eprint{hep-lat/0611014})

\bibitem{Bazavov:2009zn}
Bazavov A {\em et~al.\/} 2009 {\em Phys. Rev.\/} {\bf D80} 014504
  (\textit{Preprint} \eprint{0903.4379})

\bibitem{DeTar:2009ef}
DeTar C and Heller U~M 2009 {\em Eur. Phys. J.\/} {\bf A41} 405--437
  (\textit{Preprint} \eprint{0905.2949})

\bibitem{Fodor:2009ax}
Fodor Z and Katz S~D 2010 {\em Landolt-B\"ornstein\/} {\bf 1/23a} in press
  (\textit{Preprint} \eprint{0908.3341})

\bibitem{Fleming:2008gy}
Fleming G~T 2008 {\em PoS\/} {\bf LATTICE2008} 021 (\textit{Preprint}
  \eprint{0812.2035})

\bibitem{Pallante:2009hu}
Pallante E 2009 {\em PoS\/} {\bf LAT2009} 015 (\textit{Preprint}
  \eprint{0912.5188})

\bibitem{Beane:2008dv}
Beane S~R, Orginos K and Savage M~J 2008 {\em Int. J. Mod. Phys.\/} {\bf E17}
  1157--1218 (\textit{Preprint} \eprint{0805.4629})

\bibitem{Bulava:2009jb}
Bulava J~M {\em et~al.\/} 2009 {\em Phys. Rev.\/} {\bf D79} 034505
  (\textit{Preprint} \eprint{0901.0027})

\bibitem{Alford:2007xm}
Alford M~G, Schmitt A, Rajagopal K and Sch\"afer T 2008 {\em Rev. Mod. Phys.\/}
  {\bf 80} 1455--1515 (\textit{Preprint} \eprint{0709.4635})

\bibitem{Beane:2006gf}
Beane S~R {\em et~al.\/} (NPLQCD) 2007 {\em Nucl. Phys.\/} {\bf A794} 62--72
  (\textit{Preprint} \eprint{hep-lat/0612026})

\end{thebibliography}

\end{document}